\documentclass[authoryear,preprint,12pt]{elsarticle}

\usepackage{amssymb}
\usepackage{amsmath}
\usepackage{booktabs}
\usepackage{multirow}

\usepackage{lineno}

\journal{International Journal of Psychophysiology}

\begin{document}

\begin{frontmatter}

\title{Separating Acute Psychological Stress from Physical Exertion in Biometric Signals}

\author[label1]{Esther Bosch\corref{cor1}}
\ead{esther.bosch@dlr.de}
\ead[url]{https://orcid.org/0000-0002-6525-2650}
\cortext[cor1]{Corresponding author}
\affiliation[label1]{German Aerospace Center (DLR), Institute of Transportation Systems,\\
Braunschweig, Germany}

\begin{abstract}

Acute psychological stress occurs in a wide range of everyday contexts, including transportation, occupational settings, and physical activity, where its reliable detection could enable adaptive system responses and support human well-being. Stress produces characteristic changes in autonomic and cardiovascular physiology, yet a persistent challenge in its automated recognition is disentangling the biometric characteristics of acute psychological stress from those of concurrent physical exertion. This study examined how five physiological signals (tonic electrodermal activity, left trapezius electromyography, heart rate, heart rate variability, and respiration rate) respond to cognitive stress and physical activity, both independently and in combination. Nineteen participants completed a 2 × 3 within-subjects design in which acute psychological stress was induced via an n-back arithmetic task combined with social pressure and financial reward across three activity conditions: idle sitting, walking, and stationary cycling. Multilevel linear mixed models and repeated-measures analysis of variance were used to decompose main effects and interactions for each sensor. Tonic electrodermal activity showed a robust, additive response to both cognitive stress (r = 0.48) and physical exertion (r = 0.67), with no interaction between the two, making it the most promising candidate for stress detection during physical activity. By contrast, heart rate and trapezius electromyography were driven almost exclusively by physical exertion, with no reliable sensitivity to the n-back task. Root Mean Square of Successive Differences of heart beats was strongly suppressed by activity and showed only marginal sensitivity to cognitive load. Respiration rate was dominated by physical activity; a marginal stress
effect found in a subsidiary complete-cases analysis did not replicate
in the primary multilevel model on the full sample. These findings provide a sensor-specific hierarchy for real-world stress detection and highlight tonic electrodermal activity as the most informative channel when cognitive stress must be identified in physically active populations.
\end{abstract}


\begin{highlights}
\item Tonic electrodermal activity (EDA) showed robust sensitivity to acute 
psychological stress across all physical activity conditions, making it the 
most suitable sensor for cognitive stress detection during movement.
\item Heart rate and trapezius muscle activity (RMSA) were driven almost 
exclusively by physical exertion, with no reliable response to the stress task.
\item Respiration rate was dominated by physical activity; a marginal stress
effect found in a subsidiary complete-cases analysis did not replicate in the
primary multilevel model, and respiration rate is therefore not considered a
reliable stress indicator in the present data.
\item Heart rate variability (RMSSD) was strongly suppressed by physical 
activity, largely masking any cognitive stress signal and limiting its utility 
in ambulatory stress detection.
\item In real-world settings, decomposing the EDA signal into its stress- and
movement-related components, for instance via concurrent accelerometry, is an
important methodological challenge for field deployment.
\end{highlights}

\begin{keyword}
Acute Psychological Stress \sep Electrodermal Activity \sep Physical Exertion \sep Heart Rate Variability \sep Affective Computing \sep n-back task

\end{keyword}

\end{frontmatter}



\section{Introduction}
\label{sec1}


As modern life increasingly exposes individuals to overlapping cognitive and physical demands, the reliable detection of psychological stress in naturalistic settings has become a central challenge in affective computing \citep{schmidt2019wearable, greene2016survey}. Stress detection is inherently interdisciplinary, drawing on signal processing, machine learning, psychology, and physiology, yet the field lacks consensus on how to disentangle acute psychological stress (APS) from the physiological noise introduced by concurrent physical exertion. The difficulty is compounded in applied settings where cognitive stress and physical exertion co-occur, as is routinely the case in mobility, manual labour or emergency response work. In the field of human factors in mobility, an unresolved issue remains the difficulty of classifying affective state from biometric signals alone \citep{bosch2025travel}. Given a transport participant and their elevated biometric signals, separating  psychological arousal from physical exertion proves to be difficult. Real time stress classification systems consider this, but findings are limited \citep{lazarou2024predicting}. Understanding the interplay between physical exertion, acute psychological stress, and sensor configuration would advance real-time stress classification and detection algorithms.

This paper aims to evaluate the utility of varying sensors in ambulatory stress experiments. Affective state will be better identifiable due to characteristics produced via physical exertion. At an interdisciplinary level, this study will aim to decompose the interplay between physical exertion and acute psychological stress in biometric signals.

\section{Background}
\subsection{Physiology and Sensor Configuration for Affective Stress Recognition}
\label{sec2}
The heart is simultaneously regulated by the sympathetic nervous system (SNS) and the parasympathetic nervous system (PNS). The heart rate (HR) at any given time is identified by the overall effect of vagus nerves (PNS) which slow the rate and sympathetic nerves which accelerate it \citep{electrophysiology1996heart, shaffer2017overview}. Increased physical exertion induces sympathetic nervous system arousal to meet oxygen demand to allow for microvascular vasodilation, thereby raising HR. Similarly, psychological stress activates the sympathetic-adrenal-medullary axis, distributing catecholamines, resulting in increased activity upon HR \citep{cohen2007psychological}. In both instances, increased activity in autonomic neural pathways, blood pressure and respiratory control systems create notable shifts in heart rate variability (HRV) that are of interest when distinguishing APS from physical exertion \citep{laborde2017heart}. The Root Mean Square of Successive interbeat interval Differences (RMSSD) is a feature that indicates PNS and SNS modulation. RMSSD is the primary time-domain measure used to estimate parasympathetic modulation of heart rate variability \citep{shaffer2017overview} and is rescaled, but otherwise identical to the non-linear metric SD1, which is typically used when reflecting variability in more precise time intervals \citep{shaffer2017overview}.

Due to the increased oxygen demand in the stress response, there is a subsequent demand upon the respiratory system. The combination of respiration rate with heart activity strengthens classification of stress detection as shown by \citet{yaghoubi2025human}.

Electromyography (EMG) is a electrodiagnostic sensor that detects changes in the current of musculoskeletal functioning. The trapezius muscle has heightened myoelectrical activity in both cognitive activity and physical stress \citep{ahmed2024comprehensive}. Thus, utilising an EMG on the trapezius muscle is a common technique in stress-based experiments \citep{ahmed2024comprehensive}. It has further been discovered that the trapezius muscle activation is of particular responsiveness in individuals sensitive to anticipatory stressors; typically consisting of those with exposure to early life trauma \citep{luijcks2016impact}. The Root Mean Square Amplitude (RMSA) is a common signal characteristic that measures musculoskeletal stress \citep{wijsman2013trapezius}. Although intended to detect myoelectric activity of the trapezius muscle, RMSA is extraneously impacted by peripheral properties of surrounding muscles and privy to disruption via the sensor configuration \citep{kallenberg2006behaviour}. 

The Electrodermal Activity (EDA) is a measure of arousal formed by conductivity of the skin. Under psychophysiological arousal, skin conductance increases. Electrodermal activity is categorised into slow changing features that are unaffected by stimuli (known as tonic features), and fast event related changes (known as phasic features). Increased psychological arousal is associated with elevated skin conductance level, suggesting that conditions inducing APS would elevate skin conductance responses \citep{Boucsein2012}. The tonic and phasic components together are important features that have a proven ability to differentiate cognitive and physical exertion \citep{giannakakis2019review}.
\citet{healey2005detecting} found that especially the tonic features of EDA correlate with acute psychological stress.
Consistent with this, \citet{boettger2010heart} showed that tonic EDA increases stepwise with exercise intensity.
Palmar eccrine sweat glands are primarily activated by sympathetic arousal in response to psychological stress \citep{dawson2007electrodermal}, with physical exertion also elevating tonic EDA through increased sympathetic activity \citep{boettger2010heart}.

\subsection{Eliciting an Acute Psychological Stress Response}

Laboratory induction of acute psychological stress typically employs one of several validated paradigms, each targeting different dimensions of the stress response. The Trier Social Stress Test (TSST) is a well-validated psychosocial stress induction protocol, but requires a panel of evaluators and is poorly suited to physically active conditions \citep{kirschbaum1993trier}. The Stroop Colour-Word Task reliably elevates sympathetic arousal but is visually presented, precluding use during locomotion \citep{stroop1935studies}. The n-back task offers a viable alternative as it can be audio-delivered, allowing administration during walking and cycling without requiring visual attention, and can induce stress when combined with social evaluation, showing changes in tonic EDA and heart rate variability measured by RMSSD \citep{haucke2025development}.

\subsection{Hypotheses}
Based on the physiological mechanisms outlined above, the following hypotheses were specified prior to data collection.

\begin{enumerate}
\item For electrodermal activity, tonic EDA was expected to increase under both cognitive stress and physical exertion.
\item For heart rate, we hypothesised that both cognitive stress and physical exertion would elevate HR, as both activate the sympathetic-adrenal-medullary axis and suppress parasympathetic regulation. 
\item For heart rate variability, we predicted that RMSSD would decrease under both cognitive stress and physical exertion, reflecting increased sympathetic tone and reduced vagal modulation in both conditions, consistent with the role of RMSSD as an index of parasympathetic cardiac control.
\item For trapezius muscle activity, we expected RMSA to increase under physical exertion due to greater postural and motor demands, and to show some sensitivity to cognitive stress given evidence that the trapezius exhibits heightened myoelectrical activity under both physical and psychological load \citep{ahmed2024comprehensive}.
\item For respiration rate, we hypothesised that both cognitive stress and physical activity would independently elevate respiratory frequency, reflecting the dual role of the respiratory system in meeting metabolic demand and responding to central arousal.
\end{enumerate}

\section{Methods}

The independent variables of this experiment were stress (two levels: no stress, stress) and physical activity (three levels: Idle, cycling, walking). Both variables were manipulated within participants, yielding a $2 \times 3$ fully crossed repeated-measures design.

\subsection{Data Collection}
\label{subsec2}

Sample size was constrained by available resources, and an a priori power analysis conducted in G*Power \citep{faul2007g} was used to determine the effect size detectable under these constraints. Based on a repeated-measures analysis of variance (ANOVA) with a 2 $\times$ 3 within-subjects design, $\alpha$ = .05, and a desired power of .80, the analysis indicated that the available sample of N = 19 allows detection of effect sizes of f = 0.34, and f = 0.39 for the subsample of n = 15 with complete data. These detectable effect sizes are consistent with the medium to large effects reported in published psychophysiological stress research reporting medium to large effects for breathing rate \citep{Grassmann2016}, heart rate \citep{gu2025physiological}, heart rate variability \citep{hamidovic2020quantitative}, and RMSA on the left trapezius \citep{wijsman2013trapezius} under stress vs. no stress conditions, suggesting that the study was adequately powered to detect stress-related responses across the biosensors examined. The only sensor we could find a source to report effect sizes in stress vs no stress was tonic EDA. Most closely, \citet{doberenz2011methodological} find large effect sizes for differences in activities, but not different stress conditions. 

Participants were recruited from the institute's internal participant pool and consisted of twelve individuals that self-identified as male and six that self-identified as female (Age $39 \pm 18$ years) and one who preferred not to fill in the demographic questionnaire. Participants with any cardiovascular diseases were excluded from the study. Prior to participation, all participants were informed about the study procedures and data handling practices, gave written informed consent in accordance with the General Data Protection Regulation (GDPR), and were advised that they could withdraw at any time without providing justification. The study was reviewed and approved by the institution's ethics committee (reference no. 43/25).






\subsection{Methodology}
\label{subsec3}

Of the six experimental conditions, half were induced APS states, with the remaining half being non-APS states. For each three conditions independent of the stress state, there was varying levels of physical exertion, one of: idle while sitting, walking and stationary cycling. The conditions were randomised for each participant so that sequential effects would be counteracted. Each condition lasted eight minutes, with a five minute paced breathing exercise administered between conditions to support homeostasis. 

The stress condition was implemented in Psychopy \citep{peirce2007psychopy} as an audio based 3-back arithmetic task. Participants heard a sequence of spoken digits ranging from 100 to 900, and were asked to compare each digit to the one presented three items earlier, with digits delivered as audio recordings at 5-second intervals. In response to each spoken number, participants were instructed to press the right button of a handheld Bluetooth mouse if the number matched the one presented three items earlier, and the left button if it did not. This task has been successfully used to induce cognitive load in earlier studies \citet{walocha2025multimodal}. Additional APS was induced in the stress condition by informing participants that they would receive an extra five Euro compensation if they maintained an accuracy above 70 percent while completing the task. Social evaluative threat was introduced by having the experimenter respond to participant errors with critical facial expressions and brief negative verbal sounds, following the social-threat manipulation approach of \citet{kirschbaum1993trier}.

On the day of the experiment, participants were met at the entry check point and guided to the laboratory. Each experimental block lasted up to two hours. At the beginning, each participant signed the informed consent notice and data agreement form. Participants were then equipped with the sensors. These were a Movisens ECG Move 4 belt (1024 Hz), a Vernier Go Direct respiration belt (10Hz), a Shimmer EDA sensor (128 Hz), and a Shimmer EMG sensor (512 Hz). The 3-back arithmetic task was then explained via standardised instructions. Each participant was given as much time as they liked to practice the task. The task was administered at three-back for all participants to standardize cognitive load across the sample, as this level has been shown to reliably induce psychological stress when combined with social evaluation \citep{haucke2025development}. As manipulation check, participants rated their mental and physical demand after every condition on a scale from 1-100. After completing their six conditions, participants removed the sensors and were debriefed regarding the usage of deceit in the experiment, i.e. that every participant received five Euro extra.

\subsection{Data Processing}

We decided on the sensors based on typical sensor for stress detection \citep{giannakakis2019review}, with the limitation to use only these that would be feasible in real-world moving environments.

Skin Conductance was measured using the EDA conductance values. The signals were processed using the NeuroKit2's EDA process function \citep{makowski2021neurokit2} at a sampling rate of 128 Hz. This was further decomposed into tonic and phasic components. We continued further analysis with tonic EDA values \citep{healey2005detecting}.

The raw EMG signal was processed by subtracting calibrated values from electrode channel two with electrode channel one. Following, a fourth order Scipy Butterworth filter \citep{virtanen2020scipy}, with cutoff frequencies outside the range [0,30] was applied. The RMSA was then calculated based on \citet{pourmohammadi2020stress} by running a sliding window of 60 seconds, with a step of 30 seconds, over the filtered values. The value of RMSA at some ith index is given by equation \ref{rmsa}, where N denotes the magnitude of emissions in that 60 second range.

\begin{equation} \label{rmsa}
    RMSA[i]=\sqrt{\frac{1}{N}\sum_{k=i-N+1}^{i}(Filtered EMG[k])^2}
\end{equation}

The RMSSD of the ECG was calculated based on \citet{pourmohammadi2020stress} in a rolling frame spanning 60 seconds with a step of 30 seconds. The value of RMSSD at some ith iteration is given by equation \ref{rmssd}, where N denotes the magnitude of emissions in that 60 second range.

\begin{equation} \label{rmssd}
    RMSSD[i]=\sqrt{\frac{1}{N}\sum_{k=i-N+1}^{i}(IBI[k]-IBI[k-1]))^2}
\end{equation}

Prior to feature extraction, each signal (EDA, EMG, ECG, and respiration) was inspected and outliers outside of three standard deviations of the mean were removed. For both RMSA and RMSSD, the first and last 60-second windows were excluded from analysis, as the rolling window had not yet accumulated (or had already exceeded) sufficient data to produce valid feature estimates at the boundaries of each recording. Heart rate, RMSSD, and respiration rate were z-scored prior to analysis using each participant's own Idle no-stress condition as the reference baseline.

During data collection, technical issues with sensor recording resulted in incomplete data for three participants across one or more sensor channels, and one participant had no usable sensor data for any channel. These cases were retained in the multilevel models where partial data were available, as the mixed-model framework accommodates missing observations under a missing-at-random assumption. For the ezANOVA, which requires complete data, affected participants were excluded listwise; the resulting per-sensor sample sizes are reported in the Results.

\subsection{Data Analysis}

All statistical analyses were conducted in R (version 4.5.1; \cite{Rcore}).
To examine the effects of cognitive stress and physical activity on each physiological sensor variable, factorial repeated-measures analyses were performed separately for each of the five outcome variables: Tonic EDA, RMSA, z-scored heart rate, z- scored RMSSD, and z-scored respiration rate.

Orthogonal contrasts were specified a priori for each predictor. For stress, stress was contrasted with the no-stress condition (\textit{StressVsNone}: $-1$, $+1$). For activity, two contrasts were defined: active
conditions (cycling and walking combined) versus Idle (\textit{ActiveVsIdle}: $-2$, $+1$, $+1$), and cycling versus walking (\textit{CyclingVsWalking}: $0$, $-1$, $+1$).

Two complementary analytical approaches were applied following \citet{field2012discovering}.
First, a classical repeated-measures ANOVA was estimated using the ezANOVA
function from the ez package \citep{lawrence2016package}, with Type~III sums of
squares. This approach requires complete data, so participants with missing values in any condition were excluded listwise for this analysis only. The sphericity assumption was evaluated using Mauchly's test for all effects with three or more levels (i.e., the main effect of activity and the stress $\times$ activity interaction). Where sphericity was violated ($p < .05$), Greenhouse--Geisser (GG) corrected degrees of freedom and $p$-values are reported \citep{greenhouse1959methods}. Generalised eta-squared ($\eta^{2}_{\mathrm{G}}$) was extracted from the ezANOVA output as an omnibus
effect size measure \citep{olejnik2003generalized}.

Second, and as the primary analysis, a multilevel linear mixed model was estimated using the lme function from the nlme package \citep{pinheiro2017package},
fitted by maximum likelihood. This approach was preferred because it does
not assume sphericity, and it accommodates missing data by using all available observations
under the assumption of data missing at random. Participant was included as random factor. 

\noindent Models were built up incrementally from an intercept-only baseline, adding
the main effect of stress, then the main effect of activity, and finally the
stress $\times$ activity interaction. Each step was evaluated using a likelihood-ratio
test ($\chi^{2}$) comparing the more complex model to its predecessor \citep{field2012discovering}.
Effect sizes for individual contrasts were computed as $r$ using the formula $r = \sqrt{t^{2} / (t^{2} + df)}$ \citep{field2012discovering}, where $df$ refers to the residual degrees of freedom associated with each contrast estimate.

Post-hoc pairwise comparisons between all six stress $\times$ activity cells were conducted on the listwise-complete subsample using paired $t$-tests with Bonferroni correction. The $\alpha$ level was set at $.05$, corrected to $0.5/5 = 0.01$ for testing five different sensors.

Due to sensor failures, not all participants have complete data. Table \ref{tab:completeness} shows the available data sets per sensor.

\begin{table}[ht]
\centering
\caption{Number of participants with complete and partial data per sensor.
Complete = data present in all 6 conditions (2 stress $\times$ 3 activity);
Partial = data present in at least one but fewer than 6 conditions.}
\label{tab:completeness}
\vspace{6pt}
\begin{tabular}{lcc}
\toprule
Sensor & Complete & Partial \\
\midrule
EDA Tonic       & 15 & 0 \\
RMSA       & 10 & 1 \\
HR ($z$-score)  & 13 & 3 \\
RMSSD ($z$-score) & 10 & 4 \\
Respiration rate ($z$-score) & 15 & 3 \\
\bottomrule
\end{tabular}
\end{table}

\section{Results}
\label{sec3}

\subsection{Manipulation Check}
\begin{figure}[h]
    \centering
    \includegraphics[width=0.9\linewidth]{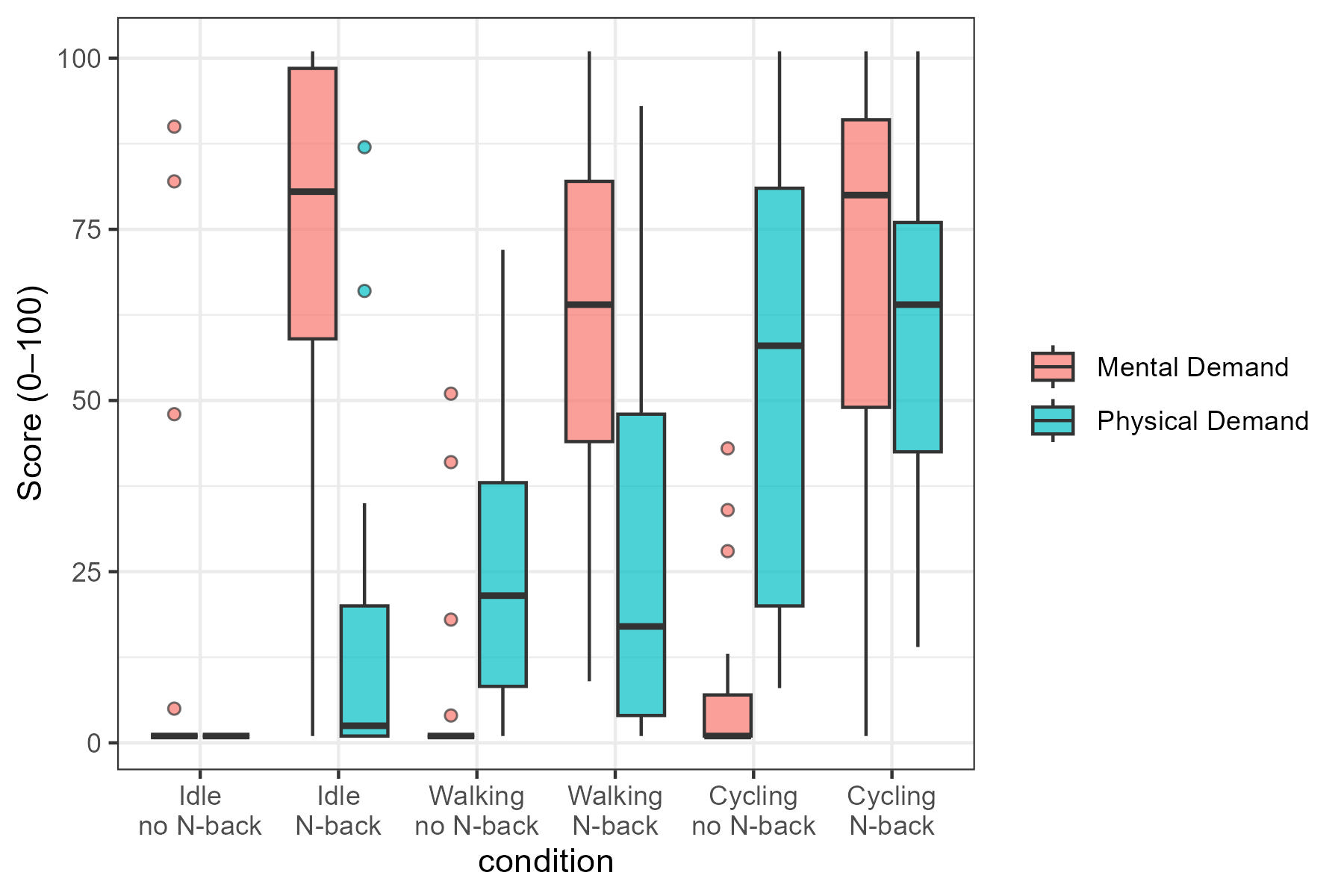}
    \caption{Mental and Physical Demand by Condition.}
    \label{mani}
\end{figure}

Mental and physical demand were reported as expected per condition, see Figure \ref{mani}.

\begin{figure}[h]
    \centering
    \includegraphics[width=0.9\linewidth]{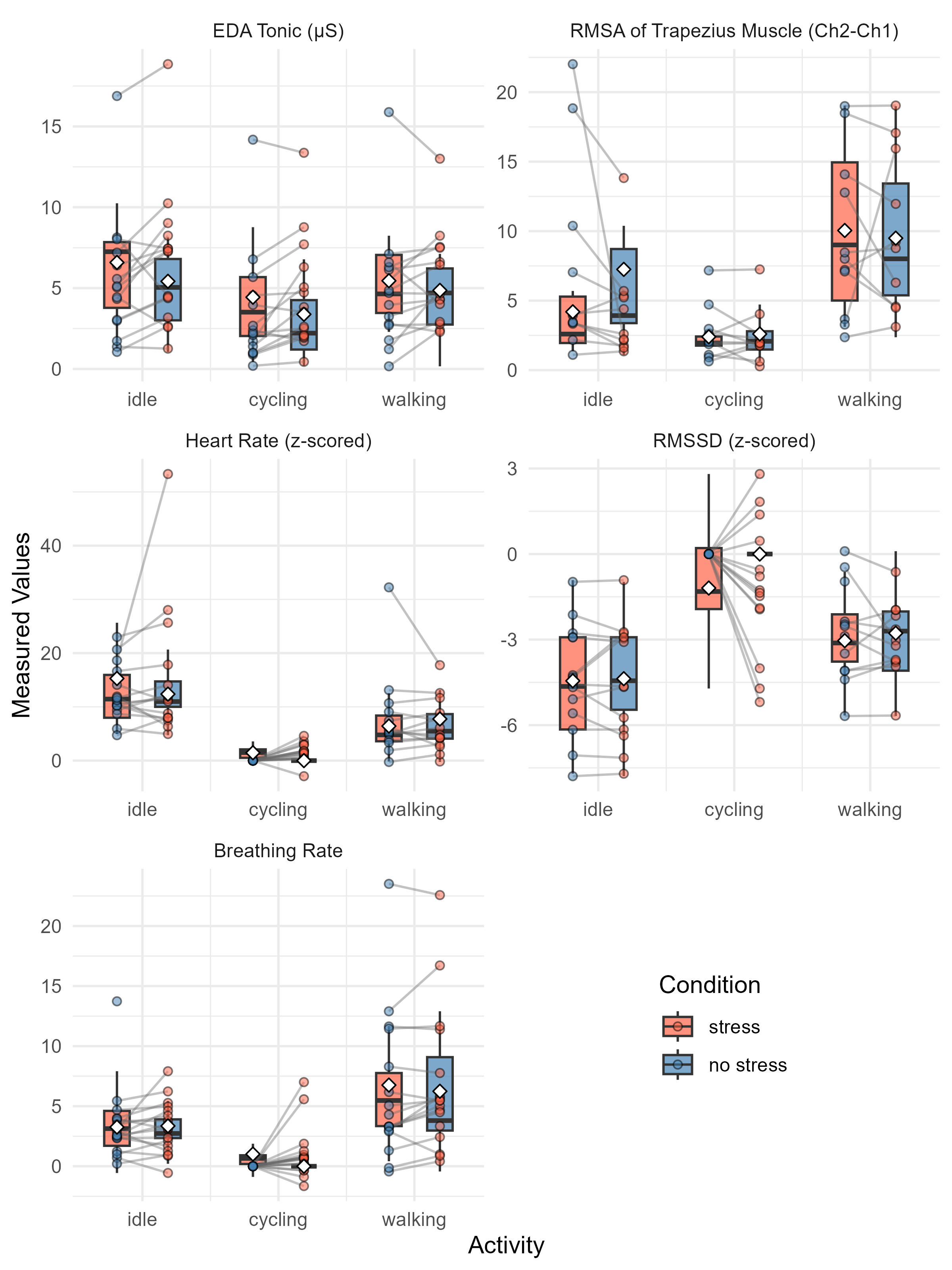}
    \caption{Physiological responses across cognitive stress x physical activity for each of the five sensor signals under no-stress (blue) and stress (red) conditions across idle, cycling, and walking. Connected lines indicate within-participant paired observations.}
    \label{boxplots}
\end{figure}

An overview of all sensor data in the different conditions can be found in Figure \ref{boxplots}. In the following, results are described per sensor.

\clearpage
\subsection{Electrodermal Activity -- Tonic Component}

Mean Tonic EDA was generally higher under active conditions
(cycling: $M = 5.42$, $SD = 3.90$ [no-stress condition]; $M = 6.59$, $SD = 4.31$ [stress condition];
walking: $M = 4.86$ [no stress condition]; $M = 5.46$ [stress condition])
relative to Idle ($M = 3.38$ [no stress condition]; $M = 4.44$ [stress condition]),
and tonic levels were consistently elevated in the stress condition across all
activity levels.

\subsubsection{Main Effect of Stress}
The multilevel model revealed a significant main effect of stress,
$\chi^{2}(1) = 7.07$, $p = .008$, $\eta^{2}_{\mathrm{G}} = .017$.
The \textit{NbackVsNone} contrast was significant,
$b = 0.47$, $t(70) = 3.23$, $p = .002$, $r = .36$,
indicating that Tonic EDA was higher during the n-back task than during
no-stress conditions.

\subsubsection{Main Effect of Activity}
Mauchly's test indicated no violation of sphericity for the activity factor ($\hat{\varepsilon}_{\mathrm{GG}} = 0.80$).
The multilevel model revealed a significant main effect of activity, $\chi^{2}(2) = 30.19$, $p < .001$, $\eta^{2}_{\mathrm{G}} = .056$.
The \textit{ActiveVsIdle} contrast was significant, $b = 0.56$, $t(70) = 5.42$, $p < .001$, $r = .54$, indicating higher tonic EDA during active compared to Idle conditions.
The \textit{CyclingVsWalking} contrast was also significant, $b = -0.42$, $t(70) = -2.38$, $p = .020$, $r = .27$,
indicating higher EDA during cycling than walking.

\subsubsection{Stress $\times$ Activity Interaction}
The multilevel model showed no significant improvement in fit when the interaction term was added, $\chi^{2}(2) = 0.77$, $p = .679$, $\eta^{2}_{\mathrm{G}} = .001$.
Neither interaction contrast reached significance (\textit{ActiveVsIdle} $\times$ \textit{NbackVsNone}: $b = -0.03$, $t(70) = -0.28$, $p = .777$, $r = .03$;
\textit{CyclingVsWalking} $\times$ \textit{NbackVsNone}: $b = -0.14$, $t(70) = -0.80$, $p = .425$, $r = .10$).
The effect of stress on Tonic EDA was therefore consistent across physical activity conditions.

\subsection{Muscle Activity -- RMS Amplitude}

Mean RMSA 60\,s was low during both Idle conditions
(none: $M = 2.60$; stress: $M = 2.40$) and higher during walking
(none: $M = 9.49$; stress: $M = 10.05$) and cycling
(none: $M = 7.25$; stress: $M = 4.20$),
reflecting a strong influence of physical movement on muscle activation.

\subsubsection{Main Effect of Stress}
The multilevel model showed no significant main effect of stress,
$\chi^{2}(1) = 0.75$, $p = .385$, $\eta^{2}_{\mathrm{G}} = .016$.
The \textit{NbackVsNone} contrast was not significant,
$b = -0.48$, $t(49) = -0.96$, $p = .343$, $r = .14$.
RMSA 60\,s did not differ significantly between stress conditions.

\subsubsection{Main Effect of Activity}
Mauchly's test indicated no violation of sphericity for the activity factor
($\hat{\varepsilon}_{\mathrm{GG}} = 0.91$).
The multilevel model revealed a significant main effect of activity,
$\chi^{2}(2) = 27.35$, $p < .001$, $\eta^{2}_{\mathrm{G}} = .306$.
The \textit{ActiveVsIdle} contrast was significant,
$b = 1.73$, $t(49) = 4.90$, $p < .001$, $r = .57$,
indicating substantially higher muscle activity during active relative to Idle conditions.
The \textit{CyclingVsWalking} contrast was also significant,
$b = 1.98$, $t(49) = 3.20$, $p = .002$, $r = .42$,
with walking showing higher RMSA than cycling.

\subsubsection{Stress $\times$ Activity Interaction}
The multilevel model showed no significant improvement with the interaction term,
$\chi^{2}(2) = 2.40$, $p = .302$, $\eta^{2}_{\mathrm{G}} = .023$.
Neither interaction contrast reached significance
(\textit{ActiveVsIdle} $\times$ \textit{NbackVsNone}:
$b = -0.19$, $t(49) = -0.54$, $p = .595$, $r = .08$;
\textit{CyclingVsWalking} $\times$ \textit{NbackVsNone}:
$b = 0.86$, $t(49) = 1.38$, $p = .173$, $r = .19$).
The effect of physical activity on RMSA was consistent regardless of cognitive stress
condition.

\subsection{z-scored Heart Rate}

Mean HR $z$-scores were substantially elevated during active conditions in both stress states (cycling none: $M = 12.38$; cycling stress: $M = 15.28$; walking none: $M = 7.75$; walking stress: $M = 6.43$).

\subsubsection{Main Effect of Stress}
The multilevel model showed no significant main effect of stress,
$\chi^{2}(1) = 0.32$, $p = .571$, $\eta^{2}_{\mathrm{G}} = .009$.
The \textit{NbackVsNone} contrast was not significant,
$b = 0.51$, $t(71) = 0.97$, $p = .335$, $r = .11$.
There was therefore no reliable effect of the stress condition on heart rate
once physical activity was accounted for.

\subsubsection{Main Effect of Activity}
Mauchly's test indicated a violation of sphericity for the activity factor,
which is why degrees of freedom were corrected using the Greenhouse--Geisser
estimate ($\hat{\varepsilon}_{\mathrm{GG}} = 0.56$).
The multilevel model revealed a significant main effect of activity,
$\chi^{2}(2) = 68.59$, $p < .001$, $\eta^{2}_{\mathrm{G}} = .408$.
The \textit{ActiveVsIdle} contrast was significant with a large effect size,
$b = 3.20$, $t(71) = 8.78$, $p < .001$, $r = .72$,
confirming that heart rate was substantially elevated during active conditions.
The \textit{CyclingVsWalking} contrast was also significant,
$b = -3.40$, $t(71) = -5.22$, $p < .001$, $r = .53$,
indicating higher heart rate during cycling than walking.

\subsubsection{Stress $\times$ Activity Interaction}
The multilevel model showed no significant improvement when the interaction
was added, $\chi^{2}(2) = 2.42$, $p = .299$, $\eta^{2}_{\mathrm{G}} = .017$.
Neither interaction contrast reached significance
(\textit{ActiveVsIdle} $\times$ \textit{NbackVsNone}:
$b = -0.11$, $t(71) = -0.31$, $p = .759$, $r = .04$;
\textit{CyclingVsWalking} $\times$ \textit{NbackVsNone}:
$b = -0.96$, $t(71) = -1.49$, $p = .141$, $r = .17$).
The pattern of heart-rate responses to physical activity did not
differ between stress conditions.

\subsection{Heart Rate Variability -- z-scored RMSSD}

RMSSD $z$-scores were markedly negative during active conditions, reflecting suppressed HRV during physical exertion (cycling none: $M = -4.37$; cycling stress: $M = -4.45$;
walking none: $M = -2.78$; walking stress: $M = -3.04$). During Idle with no stress the mean was approximately zero by construction, while the Idle stress condition showed a moderate negative shift ($M = -1.19$).

\subsubsection{Main Effect of Stress}
The multilevel model showed no significant improvement when stress was added,
$\chi^{2}(1) = 0.76$, $p = .384$, $\eta^{2}_{\mathrm{G}} = .014$.
The \textit{NbackVsNone} contrast did not reach significance,
$b = -0.20$, $t(60) = -1.29$, $p = .203$, $r = .16$.
There was therefore no significant effect of the n-back task on RMSSD.

\subsubsection{Main Effect of Activity}
Mauchly's test indicated a violation of sphericity for the activity factor;
degrees of freedom were corrected using the Greenhouse--Geisser estimate
($\hat{\varepsilon}_{\mathrm{GG}} = 0.68$).
The multilevel model revealed a significant main effect of activity,
$\chi^{2}(2) = 62.07$, $p < .001$, $\eta^{2}_{\mathrm{G}} = .584$.
The \textit{ActiveVsIdle} contrast was significant with a large effect size,
$b = -1.00$, $t(60) = -9.06$, $p < .001$, $r = .76$.
The \textit{CyclingVsWalking} contrast was also significant,
$b = 0.76$, $t(60) = 3.79$, $p < .001$, $r = .44$,
with lower RMSSD during cycling than walking.

\subsubsection{Stress $\times$ Activity Interaction}
The multilevel model showed no improvement when the interaction term
was added, $\chi^{2}(2) = 3.31$, $p = .191$, $\eta^{2}_{\mathrm{G}} = .026$.
The \textit{ActiveVsIdle} $\times$ \textit{NbackVsNone} contrast was not significant,
$b = 0.20$, $t(60) = 1.77$, $p = .082$, $r = .22$.
The \textit{CyclingVsWalking} $\times$ \textit{NbackVsNone} contrast was not
significant, $b = -0.02$, $t(60) = -0.09$, $p = .928$, $r = .01$.

\subsection{z-scored Respiration Rate}

Mean $z$-scores were elevated during active conditions (cycling none: $M = 3.36$; cycling stress: $M = 3.26$; walking none: $M = 6.24$; walking stress: $M = 6.76$) and also during the Idle stress condition ($M = 1.01$).

\subsubsection{Main Effect of Stress}
The multilevel model showed no significant main effect of stress,
$\chi^{2}(1) = 0.63$, $p = .428$, $\eta^{2}_{\mathrm{G}} = .008$.
The \textit{NbackVsNone} contrast was not significant,
$b = 0.28$, $t(81) = 0.96$, $p = .341$, $r = .11$. Note that the ezANOVA on the 15 complete-case participants yielded a significant
stress effect ($p = .029$), suggesting the effect may be fragile and sensitive
to sample composition; the more conservative lme estimate based on all 18
participants is reported here as the primary result.

\subsubsection{Main Effect of Activity}
Mauchly's test indicated a violation of sphericity for the activity factor;
degrees of freedom were corrected using the Greenhouse--Geisser estimate
($\hat{\varepsilon}_{\mathrm{GG}} = 0.55$).
The multilevel model revealed a significant main effect of activity,
$\chi^{2}(2) = 52.34$, $p < .001$, $\eta^{2}_{\mathrm{G}} = .312$.
The \textit{ActiveVsIdle} contrast was significant,
$b = 1.48$, $t(81) = 7.19$, $p < .001$, $r = .62$,
confirming higher respiration rates during active conditions.
The \textit{CyclingVsWalking} contrast was also significant,
$b = 1.52$, $t(81) = 4.17$, $p < .001$, $r = .42$,
with walking associated with higher respiration rates than cycling.

\subsubsection{Stress $\times$ Activity Interaction}
The multilevel model showed no significant improvement when the interaction
term was added, $\chi^{2}(2) = 0.40$, $p = .817$, $\eta^{2}_{\mathrm{G}} < .001$.
Neither interaction contrast reached significance
(\textit{ActiveVsIdle} $\times$ \textit{NbackVsNone}:
$b = -0.11$, $t(81) = -0.55$, $p = .585$, $r = .06$;
\textit{CyclingVsWalking} $\times$ \textit{NbackVsNone}:
$b = 0.10$, $t(81) = 0.28$, $p = .784$, $r = .03$).
The effect of physical activity on respiration rate was therefore consistent
across stress conditions.

Table \ref{tab:results_overview} shows an overview of the reported model and sensor comparisons. 

\begin{table}[ht]
\centering
\caption{Summary of multilevel model likelihood-ratio tests for all five sensor
variables. $\chi^{2}$ values are from sequential model comparisons. $\eta^{2}_{\mathrm{G}}$  is
taken from the ezANOVA on complete cases.}
\label{tab:results_overview}
\small
\begin{tabular}{llcccc}
\toprule
Sensor & Effect & $df$ & $\chi^{2}$ & $p$ & $\eta^{2}_{\mathrm{G}}$ \\
\midrule

\multirow{3}{*}{EDA Tonic}
  & Stress                   & 1 &  7.07 & .008    & .017 \\
  & Activity                 & 2 & 30.19 & $<$.001 & .056 \\
  & Stress $\times$ Activity & 2 &  0.77 & .679    & .001 \\
\midrule

\multirow{3}{*}{RMSA}
  & Stress                   & 1 &  0.75 & .385    & .016 \\
  & Activity                 & 2 & 27.35 & $<$.001 & .306 \\
  & Stress $\times$ Activity & 2 &  2.40 & .302    & .023 \\
\midrule

\multirow{3}{*}{HR ($z$-score)}
  & Stress                   & 1 &  0.32 & .571    & .009 \\
  & Activity                 & 2 & 68.59 & $<$.001 & .408 \\
  & Stress $\times$ Activity & 2 &  2.42 & .299    & .017 \\
\midrule

\multirow{3}{*}{RMSSD ($z$-score)}
  & Stress                   & 1 &  0.76 & .384    & .014 \\
  & Activity                 & 2 & 62.07 & $<$.001 & .584 \\
  & Stress $\times$ Activity & 2 &  3.31 & .191    & .026 \\
\midrule

\multirow{3}{*}{Respiration rate ($z$-score)}
  & Stress                   & 1 &  0.63 & .428    & .008 \\
  & Activity                 & 2 & 52.34 & $<$.001 & .312 \\
  & Stress $\times$ Activity & 2 &  0.40 & .817    & .000 \\

\bottomrule
\end{tabular}
\end{table}

\clearpage

\section{Discussion}
\label{sec4}

This study researched how stress and physical activity affect different physiological sensor signals in a 2~$\times$~3 within-subjects design. The results show that tonic EDA was sensitive to both stress and physical activity, whereas respiration rate, heart rate, HRV (RMSSD) and muscle activity (RMSA)  mainly changed due to physical activity. These findings have implications for the selection of sensors in real-world stress detection such as everyday travel.  

Tonic EDA was the only measure to show a robust and significant main effect between stress and no stress ($r_\text{StressVsNostress} = .48$), together with a strong effect of physical activity ($r_\text{ActiveVsIdle} = .67$). The absence of a stress~$\times$~activity interaction indicates that the sympathetic response to the stress condition was additive with physical activity. This finding is in line with the previously described physiology of tonic skin conductance: because eccrine sweat glands are innervated exclusively by the sympathetic nervous system and have no parasympathetic input, Tonic EDA shows a slow-changing level of sympathetic arousal that adds up both from stress and from physical activity~\citep{Boucsein2012, PosadaQuintero2018}. Previous research has shown that cognitive stressors including mental arithmetic, loss of money, driving and Stroop tasks elevate tonic skin conductance~\citep{giakoumis2012using, reinhardt2012salivary, healey2005detecting}, and our results extend this to situations where participants are cycling or walking during the stress induction. The increase in Tonic EDA during active conditions, with cycling producing somewhat higher values than walking ($r_\text{CyclingVsWalking} = .37$), is in line with previous research showing that exercise intensity progressively elevates tonic skin conductance through a combination of thermoregulatory sweating and heightened central sympathetic drive \citep{PosadaQuintero2018}. 

Heart rate was strongly influenced by physical activity
($r_\text{ActiveVsIdle} = .81$), and was higher during cycling than walking
($r_\text{CyclingVsWalking} = .64$).
It showed no significant effect of the stress condition
($\chi^{2}(1) = 0.32$, $p = .571$, $r_\text{NbackVsNone} = .11$).
These results are consistent with the metabolic demands of cycling and
walking elevating heart rate through autonomic and mechanical mechanisms
that outweigh the sympathetic activation produced by stress
\citep{nobrega2014neural, michael2017cardiac}.
Similarly, research that combined an n-back task with exercise has found
that cognitive load during self-paced cycling does not significantly alter
heart rate relative to exercise without a concurrent cognitive task
\citep{holgado2019no}.
This suggests that the metabolic signal overwhelms the cognitive stress
signal in the heart rate measure.
Trapezius muscle activity was also insensitive to stress and driven
exclusively by physical activity
($r_\text{ActiveVsIdle} = .66$, $r_\text{CyclingVsWalking} = .49$).
Walking resulted in higher RMSA than cycling, which is consistent with
the whole-body postural demands and upper-limb movement of walking.
These results show that RMSA in the trapezius reflects the postural and
motor demands of the activity rather than stress, at least within the
range of stress induced by the n-back task combined with money loss and
social evaluation.
Together, the HR and RMSA results indicate that these two measures are
poorly suited as stress indicators in studies that include physical activity.

Heart rate variability as measured by RMSSD was much lower during physical
activity than sitting, and also lower during cycling than walking
($r_\text{ActiveVsIdle} = .82$, $r_\text{CyclingVsWalking} = .53$).
This is consistent with lower cardiac parasympathetic activity during
exercise as sympathetic tone increases \citep{michael2017cardiac}.
The effect of physical activity on RMSSD was stronger than the effect
of stress: neither the main effect of stress nor the \textit{NbackVsNone}
contrast in the multilevel model reached significance
($\chi^{2}(1) = 0.76$, $p = .384$, $r = .16$).
This stands in contrast to previous literature, where RMSSD is frequently
reported to be sensitive to psychological stress
\citep{immanuel2023heart, giannakakis2019review}.
This is likely due to the fact that existing studies most commonly measure
HRV at rest, where physical activity does not compete with the stress signal.
The interaction between activity and stress did not reach significance
($\chi^{2}(2) = 3.31$, $p = .191$), and neither interaction contrast was
significant, though the \textit{ActiveVsIdle} $\times$ \textit{NbackVsNone}
contrast showed a non-significant trend ($p = .082$, $r = .22$).
This pattern could tentatively suggest that the HRV suppression during
active conditions was partly attenuated under stress relative to the
no-stress condition, possibly reflecting a ceiling effect in sympathetic
activation: once physical activity has already strongly suppressed
parasympathetic tone, any additional sympathetic drive from stress may
produce only a small and inconsistent further reduction in RMSSD.


Respiration rate changed significantly with physical activity
($r_\text{ActiveVsIdle} = .64$; $r_\text{CyclingVsWalking} = .41$),
with no interaction between stress and activity.
The primary analysis using the multilevel model on all 18 participants
showed no significant main effect of stress
($\chi^{2}(1) = 0.63$, $p = .428$, $r = .11$).
A supplementary ezANOVA on the 15 complete-case participants yielded a
marginal stress effect ($p = .029$), suggesting a possible but fragile
sensitivity of respiration rate to cognitive load that did not survive
in the more conservative and complete analysis.
This discrepancy is likely attributable to the difference in sample size
between the two analyses rather than a true effect, and we therefore do
not interpret respiration rate as a reliable stress indicator in the
present data.
This is somewhat at odds with a systematic review by \citet{Grassmann2016},
who concluded that mentally demanding tasks reliably increase breathing
frequency.
The discrepancy may be due to the high variability in breathing patterns
induced by physical activity masking any stress-related signal.
That walking produced higher respiration rates than cycling ($r = .41$)
is somewhat surprising given that cycling typically yields higher heart
rates in the present data.
This could be due to differences in the intensity at which participants
paced themselves.
Future work could benefit from objectively controlling exercise intensity
(e.g., via fixed heart rate targets) across activity conditions to
disentangle pacing-related from modality-specific respiratory effects. 

Overall, the five measures differ in their suitability for detecting stress
during physical activity.
Tonic EDA was the strongest and most robust stress indicator across all
activity conditions.
RMSSD showed only marginal and inconsistent sensitivity to stress, largely
overridden by physical activity-related parasympathetic withdrawal.
Heart rate and trapezius muscle activity showed no sensitivity to the stress
condition during movement.
Respiration rate showed a marginal stress effect in the complete-cases
analysis only, and is therefore not considered a reliable stress indicator
in the present data.
These results have practical implications for the design of real-world
physiological monitoring systems, including transportation contexts where
commuters may walk, cycle, or sit~\citep{bosch2025travel}.
A sensor suite relying on heart rate alone would not be able to detect
stress during active commuting modes.
Tonic EDA, despite its sensitivity to thermoregulatory sweating during
exercise, is a more promising measure for cognitive stress detection during
physical activity, provided that the activity-related baseline elevation
is accounted for analytically.
Whether combining tonic EDA with additional sensors such as respiration
rate offers further robustness remains an open question that warrants
investigation in larger samples. 

A practical challenge that follows from the present findings concerns the
decomposition of the EDA signal in real-world settings: when tonic EDA is
elevated during active commuting, it is not immediately clear whether the
increase reflects cognitive stress, physical exertion, or both.
One promising approach is the concurrent measurement of physical activity
via accelerometry, which would allow the activity-related EDA component to
be estimated and statistically controlled, leaving a residual signal more
specific to stress \citep{PosadaQuintero2018}.
Alternatively, individual calibration sessions, in which each
participant's EDA response to varying movement intensities is recorded
under stress-free conditions, could provide a person-specific
activity--EDA function that serves as a baseline for subtraction in
subsequent recordings.
Multimodal modelling approaches that combine EDA with other signals (e.g.\
heart rate, respiration rate, accelerometry) may further improve the
separation of stress- and movement-related variance, since physical
activity affects all sensors broadly whereas cognitive stress produces
a more specific pattern of autonomic activation.
Developing and validating such decomposition methods in naturalistic
settings is an important direction for future research.

Several limitations should be mentioned regarding this research. First, the sample size was small and, depending on the sensor, complete data was only obtained from $N = 10$ to $N = 15$ participants. This reduces the statistical power especially for detecting interaction effects in the calculated ANOVAs. The multilevel models retained more participants by handling missing data under a missing-at-random assumption, but replication with larger samples is needed. Second, the n-back task with social evaluation and money loss is a specific type of social and cognitive stress; other stressor types (e.g., time pressure or emotional arousal) may lead to different patterns, especially for HRV and heart rate~\citep{Kreibig2010}. Third, exercise intensity was not clearly set for walking and cycling, which means that some of the within-sensor variance may be due to differences in self-selected pace rather than the experimental manipulations. Future studies should consider standardising physical load (e.g., by target heart rate zones) to improve comparability across activity conditions and participant samples. Finally, all data were collected in a laboratory setting; whether the additive stress structure observed here generalises to naturalistic ambulatory contexts, where cognitive demands, physical intensity, and environmental stimuli co-vary more freely, remains an open question for field-based research.

\section{Conclusion}
\label{sec5}
This study examined five physiological measures to detect acute psychological stress under varying levels of physical activity. Tonic EDA was the only measure to show a robust and consistent stress signal
across all activity conditions, responding to both cognitive stress and
physical exertion in an additive, non-interacting manner.
Heart rate and trapezius muscle activity were dominated almost entirely by
physical activity, with no reliable sensitivity to the n-back task, and are
therefore unsuitable stress measures in moving environments.
RMSSD was similarly dominated by physical activity, with only a marginal and
inconsistent stress signal that is unlikely to be practically useful in
physically active contexts.
Respiration rate showed a marginal stress effect in a subsidiary analysis on
complete cases only, which did not replicate in the primary multilevel model
on the full sample; it is therefore also not considered a reliable stress
indicator here.

For the measurement of stress in physically active populations, such as active commuting, occupational tasks, or emergency response work, this means that measurements relying on heart rate and heart rate variability only (as often the case when smartwatches are used) would systematically fail to capture cognitive stress during physical activity. Tonic EDA, by contrast, preserves a detectable and consistent stress signal across all activity conditions, provided that the activity-related baseline elevation is accounted for. The combination of tonic EDA and respiration rate offers a particularly promising multimodal configuration, as both signals showed additive sensitivity to the two stressor types without mutual interference.

The present results should be interpreted in light of several constraints. The sample was small, and the stressor targeted working memory supplemented by social-evaluative pressure and financial incentive, which may not generalise to other forms of psychological stress. Future work should examine whether the additive stress structure observed here holds under ecologically valid conditions, including naturalistic commuting environments, emotionally arousing stressors, and a wider range of physical intensities. Extending this sensor hierarchy to real-world ambulatory settings remains an important next step toward robust, context-aware stress detection. A key methodological challenge for such field deployments will be the
decomposition of the EDA signal into its stress- and movement-related
components, for instance through concurrent accelerometry or
individual activity calibration, so that cognitive stress can be
reliably quantified in the presence of physical activity.

\section{Declaration of generative AI and AI-assisted technologies in the writing process.}

Statement: During the preparation of this work, the authors used Claude Sonnet 4.6 (Version April 2026) to support spell-checking and refine wording. After using this tool, the authors reviewed and edited the content as needed and take full responsibility for the content of the published article.

\section{Declaration of Interest.}
The authors declare no conflict of interest.

\section{Acknowledgements.}

CRediT roles: all EB.

Funding: This work was supported by the DLR (German Aerospace Center) project MoDa (Models and Data for Future Mobility - Supporting Services). 


\bibliographystyle{elsarticle-harv} 
\bibliography{02_references}

\end{document}